# Spin-Hall effect and circular birefringence of a uniaxial crystal plate

Konstantin Y. Bliokh,[1,2] C. T. Samlan,[3] Chandravati Prajapati,[3]
Graciana Puentes,[4] Nirmal K. Viswanathan,[3] and Franco Nori[1,5]

[1]*Center for Emergent Matter Science, RIKEN, Wako-shi, Saitama 351-0198, Japan*
[2]*Nonlinear Physics Centre, RSPhysE, The Australian National University, Canberra, ACT 0200, Australia*
[3]*School of Physics, University of Hyderabad, Hyderabad 500046, India*
[4]*Departamento de Fisica, Facultad de Ciencias Exactas y Naturales, Pabellon 1, Ciudad Universitaria, 1428 Buenos Aires, Argentina*
[5]*Physics Department, University of Michigan, Ann Arbor, MI 48109-1040, USA*

The linear birefringence of uniaxial crystal plates is known since the 17[th] century, and it is widely used in numerous optical setups and devices. Here we demonstrate, both theoretically and experimentally, a fine lateral *circular* birefringence of such crystal plates. This effect is a novel example of the spin-Hall effect of light, i.e., a transverse spin-dependent shift of the paraxial light beam transmitted through the plate. The well-known linear birefringence and the new circular birefringence form an interesting analogy with the Goos–Hänchen and Imbert–Fedorov beam shifts that appear in the light reflection at a dielectric interface. We report the experimental observation of the effect in a remarkably simple system of a tilted half-wave plate and polarizers using polarimetric and quantum-weak-measurement techniques for the beam-shift measurements. In view of great recent interest in spin-orbit interaction phenomena, our results could find applications in modern polarization optics and nano-photonics.

## 1. Introduction

*Spin-orbit interactions* (SOIs) of light have attracted ever-growing interest during the past decade [1,2]. Because of their fundamental origin and generic character, SOI phenomena have become inherent in the areas of nano-optics, singular optics, photonics, and metamaterials. Indeed, SOIs manifest themselves in the most basic optical processes – propagation, reflection, diffraction, scattering, focusing, etc. – as soon as these processes are carefully considered at subwavelength scales. In this work, we describe a novel spin-orbit phenomenon, which occurs in a very simple and thoroughly studied optical system, namely, a thin uniaxial-crystal plate.

The majority of SOI effects originate from the space- or wavevector-variant geometric phases, which result in the spin-dependent redistribution of the light intensity [1]. First, when the system possesses cylindrical symmetry with respect to the $z$-axis, SOIs produce *spin-to-orbital angular momentum conversion*, i.e., generation of a spin-dependent vortex in the $z$-propagating light [3–14]. Second, if the cylindrical symmetry is broken, say, along the $x$-direction, SOIs bring about the *spin-Hall effect* of light, i.e., a spin-dependent transverse $y$-shift of the light intensity [11–24]. An example of the latter effect is the so-called transverse Imbert–Fedorov beam shift, which occurs when a paraxial optical beam is reflected or refracted at a plane interface [20–24].

The two main factors, which typically induce the SOI effects, are: (i) the medium *inhomogeneity*, which changes the direction of propagation of light and (ii) the *anisotropy*, which induces the phase difference between two polarization components of light [1]. Table I summarizes the above two types of SOI effects in inhomogeneous (but isotropic) and anisotropic (but homogeneous) systems. For instance, the radial *inhomogeneity* in the cylindrically-symmetric focusing or scattering systems results in the spin-to-orbital angular momentum conversion [5–14], and a very similar effect appears in the paraxial light propagation along the optical axis in cylindrically-symmetric *anisotropic* media [25,26]. As we mentioned above, the simplest example of the spin-Hall effect occurs in the reflection or refraction of light at a sharp *inhomogeneity* of an isotropic optical interface [20–24]. Then, what is the counterpart of this phenomenon for the paraxial light propagating in an *anisotropic* medium?

In this paper we demonstrate, both theoretically and experimentally, that the spin-Hall effect of light and the transverse spin-dependent beam shift appears in the light transmission through a *uniaxial crystal plate* (such as wave plates routinely used in optics) with a *tilted* anisotropy axis. This new type of spin-Hall effect is quite surprising for traditional optics, because it implies a weak *circular birefringence* of a uniaxial crystal plate. Indeed, the linear birefringence of a calcite plate is known since the 17[th] century [27], while here we demonstrate a circular birefringence of such a crystal in the orthogonal direction. Notably, the well-known linear birefringence and novel circular birefringence exhibit a close similarity with the Goos–Hänchen (GH) and Imbert–Fedorov (IF) beam shifts in the reflection/refraction of light at isotropic interfaces. We provide experimental measurements of this effect using a standard half-wave plate tilted with respect to the laser beam.

The spin-Hall shift in the transmission of light through a uniaxial-crystal plate has the same order of magnitude as the IF beam shifts, i.e., a fraction of the wavelength. We use polarimetric techniques [28] to characterize the circular-polarization splitting and shifts of the transmitted beam. We also amplify the effect to the beam-width scale, using the "quantum weak measurement" approach [29–32]. This method was previously employed to amplify the usual linear birefringence of uniaxial crystals [33,34] and IF beam shifts at isotropic interfaces [22,35–41].

|  | **Spin-to-orbital AM conversion** | **Spin-Hall effect for paraxial beams** |
|---|---|---|
| **Inhomogeneity** | Focusing/scattering in cylindrically-symmetric systems [5–14] | Reflection/refraction at a plane interface (IF shift) [20–24] |
| **Anisotropy** | Propagation along the optical axis in a uniaxial crystal [25,26] | Transmission through a uniaxial-crystal plate with tilted optical axis |

**Table I.** Basic spin-orbit interaction effects, spin-to-orbital angular momentum conversion and spin-Hall effect, in inhomogeneous isotropic and anisotropic homogeneous systems. The present work completes this table by the highlighted cell.

## 2. Gaussian beam transmitted through a uniaxial crystal plate

To begin with, we consider polarized paraxial Gaussian beams propagating along the $z$-axis in free space. The beam represents a superposition of multiple plane waves (spatial Fourier harmonics) with close wave vectors

$$\mathbf{k} = k_z \overline{\mathbf{z}} + k_x \overline{\mathbf{x}} + k_y \overline{\mathbf{y}} \simeq k\left(1 - \frac{\Theta^2}{2}\right)\overline{\mathbf{z}} + k\Theta_x \overline{\mathbf{x}} + k\Theta_y \overline{\mathbf{y}}, \qquad (1)$$

where $k$ is the wave number, $\overline{\mathbf{x}}, \overline{\mathbf{y}}, \overline{\mathbf{z}}$ are the unit vectors of the corresponding axes, while $\boldsymbol{\Theta} = (\Theta_x, \Theta_y)$, $\Theta^2 = \Theta_x^2 + \Theta_y^2 \ll 1$, are small angles of the wave vector with respect to the $z$-axis in the $(x,z)$ and $(y,z)$ planes (see Fig. 1). The Fourier (momentum) representation of the transverse electric field of the Gaussian beam can be written as [21,23,24]

$$\tilde{\mathbf{E}}_\perp(\boldsymbol{\Theta}) \propto \begin{pmatrix} \alpha \\ \beta \end{pmatrix} \exp\left[-(kw_0)^2 \frac{\Theta^2}{4}\right], \qquad (2)$$

where $\begin{pmatrix} \alpha \\ \beta \end{pmatrix}$ is the normalized Jones vector of the wave polarization in the $(x,y)$ basis, $|\alpha|^2 + |\beta|^2 = 1$, and $w_0$ is the beam waist. Performing the Fourier transform of Eq. (2), we obtain the transverse beam field in the real-space representation:

$$\mathbf{E}_\perp(\mathbf{R}) \propto \int \tilde{\mathbf{E}}_\perp(\mathbf{\Theta}) e^{i\mathbf{k}\cdot\mathbf{r}} d^2\mathbf{\Theta} \propto \begin{pmatrix} \alpha \\ \beta \end{pmatrix} \exp\left[-\frac{\mathbf{R}^2}{w_0^2}\right]. \quad (3)$$

Here $d^2\mathbf{\Theta} = d\Theta_x d\Theta_y$, $\mathbf{R} = (x,y)$, $\mathbf{R}^2 = x^2 + y^2$, is the transverse radius-vector, and for simplicity we calculated the beam at the waist plane $\mathbf{E}_\perp(\mathbf{R}) \equiv \mathbf{E}_\perp(\mathbf{r})|_{z=0}$. The transition between the momentum (2) and real-space (3) representations of the beam can be realized using the stationary-phase asymptotic at the stationary point $\mathbf{\Theta}^s(\mathbf{R}) = i\mathbf{R}/z_R$: $\mathbf{E}_\perp(\mathbf{R}) = \tilde{\mathbf{E}}_\perp[\mathbf{\Theta}^s(\mathbf{R})]$, where $z_R = kw_0^2/2$ is the Rayleigh range.

Now, let us consider the transmission of the Gaussian beam (1)–(3) through a thin uniaxial-crystal (e.g., calcite or quartz) plate. The beam still propagates along the $z$-axis, whereas the anisotropy axis of the plate lies in the $(x,z)$ plane at the angle $-\vartheta$ with respect to the $z$-axis, Fig. 1. It is well-known that the crystal plate induces linear birefringence between the *ordinary* (*o*) and *extraordinary* (*e*) polarization modes, which propagate with slightly different phase velocities. For the central plane wave in the beam, $\mathbf{\Theta} = 0$, the extraordinary and ordinary modes correspond to the $x$- and $y$- linear polarizations: $\begin{pmatrix} \alpha \\ \beta \end{pmatrix}_e = \begin{pmatrix} 1 \\ 0 \end{pmatrix}$ and $\begin{pmatrix} \alpha \\ \beta \end{pmatrix}_o = \begin{pmatrix} 0 \\ 1 \end{pmatrix}$ (these can also be called TM and TE modes, respectively). Thus, the action of the plate on the central plane wave can be characterized by the following Jones equation and matrix:

$$\tilde{\mathbf{E}}'_\perp(\mathbf{0}) = \hat{M}_0 \tilde{\mathbf{E}}_\perp(\mathbf{0}), \quad \hat{M}_0 = \begin{pmatrix} e^{-i\Phi_0/2} & 0 \\ 0 & e^{i\Phi_0/2} \end{pmatrix}, \quad (4)$$

where $\Phi_0(\vartheta)$ is the phase difference between the *o* and *e* modes, which is acquired upon the propagation in the plate, the prime indicates the field of the transmitted wave, and we ignore the common phase factor.

Importantly, the zero-order transmission Jones matrix (4) is exact only for the central plane wave in the beam, $\mathbf{\Theta} = 0$. Fourier components with $\mathbf{\Theta} \neq 0$ propagate in slightly different directions and, hence, are described by slightly different Jones matrices. First, the waves with the in-plane deflection $\Theta_x \neq 0$ propagate at *angles* $\theta \simeq \vartheta + \Theta_x$ *to the anisotropy axis* (Fig. 1b). This slightly modifies the phase difference between the *o* and *e* polarizations of such waves: $\Phi(\theta) \simeq \Phi_0 + \frac{d\Phi_0}{d\vartheta}\Theta_x$. Second, the waves with the out-of-plane deflection $\Theta_y \neq 0$ propagate (in the linear approximation in $\Theta_y$) at the same angle to the anisotropy axis but in slightly different *planes of propagation*, which are *rotated* about the anisotropy axis by the azimuthal angle $\phi \simeq \Theta_y / \sin\vartheta$ (Fig. 1c). Such rotation induces additional geometric phases for circularly-polarized plane waves with $\Theta_y \neq 0$, i.e., effects of *spin-orbit interaction* of light (see the detailed descriptions in [1,24]). The above corrections modify the reflection Jones matrix (4), resulting in the $\mathbf{\Theta}$-dependent terms:

$$\tilde{\mathbf{E}}'_\perp(\mathbf{\Theta}) = \hat{\tilde{M}}(\mathbf{\Theta}) \tilde{\mathbf{E}}_\perp(\mathbf{\Theta}), \quad \hat{\tilde{M}} \simeq \begin{pmatrix} e^{-i\Phi_0/2}(1+\Theta_x \mathcal{X}_e) & e^{-i\Phi_0/2}\Theta_y \mathcal{Y}_e \\ -e^{i\Phi_0/2}\Theta_y \mathcal{Y}_o & e^{i\Phi_0/2}(1+\Theta_x \mathcal{X}_o) \end{pmatrix}, \quad (5)$$

where

$$\mathcal{X}_{e,o} = \mp \frac{i}{2}\frac{d\Phi_0}{d\vartheta}, \quad \mathcal{Y}_{e,o} = \left[1 - \exp(\pm i\Phi_0)\right]\cot\vartheta. \tag{6}$$

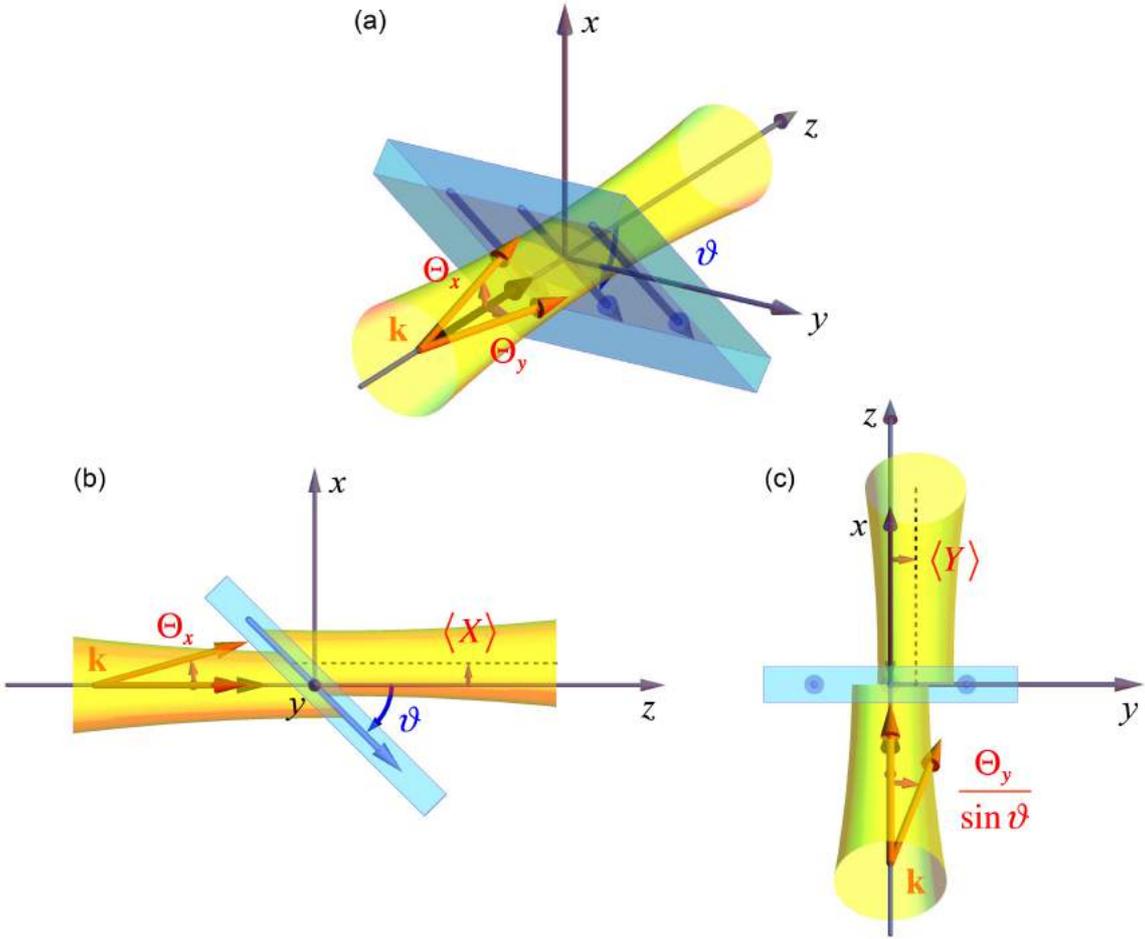

**Figure 1.** Schematics of the transmission of a paraxial beam through a tilted uniaxial-crystal plate. **(a)** General 3D geometry of the problem with the angle $\vartheta$ between the anisotropy axis of the plate and the beam axis $z$. The small angles $\Theta = (\Theta_x, \Theta_y)$ determine the directions of the wave vectors **k** in the incident beam. **(b)** The in-plane $\Theta_x$-deflections of the wave vectors change the angle between **k** and the anisotropy axis and result in the well-known birefringence shift $\langle X \rangle$, analogous to the Goos–Hänchen shift [24]. **(c)** The view along the anisotropy axis of the crystal is shown. The transverse $\Theta_y$-deflections of the wave vectors rotate the corresponding planes of the wave propagation with respect to the anisotropy axis by the angle $\phi \simeq \Theta_y / \sin\vartheta$. This causes a new helicity-dependent transverse shift $\langle Y \rangle$, i.e., a circular birefringence or spin-Hall effect similar to the Imbert–Fedorov shift [20–24].

Remarkably, we notice a one-to-one correspondence between the effective $\Theta$-dependent Jones matrix (5) and (6) and a similar matrix for the total internal reflection of the beam at a dielectric interface [24]. In this manner, the $e$ and $o$ polarization modes of the crystal correspond to the TM ($p$) and TE ($s$) modes of the interface, and the phase difference $\Phi_0$ corresponds to the difference between the phases of the Fresnel reflection coefficients for the $p$ and $s$ modes in the total internal reflection. (Note that there are some inessential differences in signs in anisotropic-plate and total-internal-reflection equations, which appear because of the difference between the transmission and reflection geometries.) In the beam reflection from an interface, the terms proportional to $\mathcal{X}_{p,s}$ describe the in-plane GH beam shift [24], while the $\mathcal{Y}_{p,s}$ terms are responsible for the transverse IF shift or spin-Hall effect of light [20–24]. Therefore, the beam transmission through a uniaxial crystal plate must exhibit similar shifts. In this manner, the GH-like shifts described by the $\mathcal{X}_{e,o}$ correspond to the usual linear birefringence between the ordinary and extraordinary rays, while the shifts described by $\mathcal{Y}_{e,o}$ correspond to a new type of the *spin Hall effect of light* and effective *circular* transverse birefringence of a uniaxial plate.

The $\Theta$-dependent Jones matrix (5) describes the transformation of the paraxial beam field in the momentum representation. To write this field transformation in the coordinate representation, we make the Fourier transform of Eq. (5), $\hat{M}(\mathbf{R}) = \hat{\tilde{M}}[\Theta^s(\mathbf{R})]$:

$$\mathbf{E}'_\perp(\mathbf{R}) = \hat{M}(\mathbf{R})\mathbf{E}_\perp(\mathbf{R}), \quad \hat{M} \simeq \begin{pmatrix} e^{-i\Phi_0/2}\left(1+i\dfrac{x}{z_R}\mathcal{X}_e\right) & ie^{-i\Phi_0/2}\dfrac{y}{z_R}\mathcal{Y}_e \\ -ie^{i\Phi_0/2}\dfrac{y}{z_R}\mathcal{Y}_o & e^{i\Phi_0/2}\left(1+i\dfrac{x}{z_R}\mathcal{X}_o\right) \end{pmatrix}. \quad (7)$$

The real-space Jones matrix (7) contains $\mathbf{R}$-dependent terms, which describe an *inhomogeneous polarization distribution* in the cross-section of the transmitted beam. Most importantly, even for the *e* and *o* polarizations of the incident beam, the transmitted field exhibits an inhomogeneous distribution of *elliptical* polarizations due to the $y$-dependent terms in $\hat{M}(\mathbf{R})$. For example, taking the *e*-polarized incident beam (3) with $\begin{pmatrix}\alpha\\\beta\end{pmatrix}=\begin{pmatrix}1\\0\end{pmatrix}$, the transmitted beam field (7) yields

$$\mathbf{E}'_\perp(\mathbf{R}) \propto \begin{pmatrix} 1+i\dfrac{x}{z_R}\mathcal{X}_e \\ -ie^{i\Phi_0}\dfrac{y}{z_R}\mathcal{Y}_o \end{pmatrix} \exp\left[-\dfrac{\mathbf{R}^2}{w_0^2}\right], \quad (8)$$

Since $-ie^{i\Phi_0}\mathcal{Y}_o = [\sin\Phi_0 + i(1-\cos\Phi_0)]\cot\vartheta$, for $\Phi_0 \neq 0 \bmod 2\pi$ the beam polarization acquires weak a right-hand ellipticity at $y>0$ and a left-hand ellipticity at $y<0$ in the $x=0$ cross-section, as shown in Fig. 2. This signifies the transverse $y$-splitting of the right-hand and left-hand circular-polarization components in the beam [13,20], i.e., the spin Hall effect of light. This splitting can be measured via direct polarimetric methods (see Fig. 6 below) or detected by placing a crossed *o*- ($y$-axis) polarizer after the crystal plate. Such polarizer cuts the $y$-component of the transmitted field and produces a two-hump Hermitte–Gaussian intensity distribution (see Fig. 7 below) [20]:

$$E'_{\perp y}(\mathbf{R}) \propto \dfrac{y}{z_R}\mathcal{Y}_o \exp\left[-\dfrac{\mathbf{R}^2}{w_0^2}\right]. \quad (9)$$

## 3. Beam shifts and their amplification via quantum weak measurements

We are now at the position to calculate the beam shifts induced by the $\mathcal{X}_{e,o}$ and $\mathcal{Y}_{e,o}$ terms in Eqs. (5)–(7). The beam shifts can be determined straightforwardly by calculating the centroid of the intensity distribution of the transmitted field $\mathbf{E}'_\perp(\mathbf{R})$ [20,21,23,24]. However, since we will also use a method of quantum weak measurements [29–32] to amplify and detect the shifts, we will follow the general quantum-mechanical-like formalism developed in [38–40].

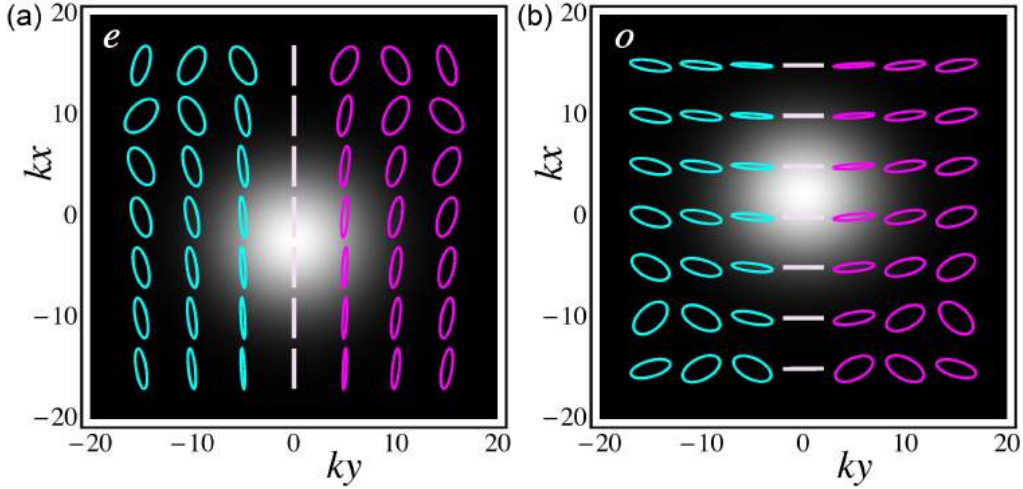

**Figure 2.** Distribution of the polarization in the extraordinary **(a)** and ordinary **(b)** Gaussian beams transmitted through a uniaxial crystal plate. Right-hand and left-hand polarization ellipses are shown in magenta and cyan, respectively. The background grayscale distributions show the Gaussian intensities of the beams, $x$-shifted due to ordinary linear birefringence. Transverse $y$-dependent separations of opposite helicities and tilts of the polarization ellipses indicate the $\sigma$- and $\chi$-dependent transverse birefringence, described by the $\mathcal{Y}_{e,o}$ and $\langle Y \rangle$ terms in Eqs. (5)–(9). The parameters used here are: $kw_0 = 10$, $\vartheta = \pi/4$, $\Phi_0 = 2\pi/3$, and $d\Phi_0/d\vartheta = -5$.

We first write the momentum-representation Jones matrix (5) as [40] $\hat{\tilde{M}} = \left(1 - ik\mathbf{\Theta}\cdot\hat{\mathbf{R}}\right)\hat{M}_0$, and introduce the matrix operators $\hat{\mathbf{R}} = (\hat{X}, \hat{Y})$:

$$\hat{X} = ik^{-1}\begin{pmatrix} \mathcal{X}_e & 0 \\ 0 & \mathcal{X}_o \end{pmatrix}, \quad \hat{Y} = ik^{-1}\begin{pmatrix} 0 & e^{-i\Phi_0}\mathcal{Y}_e \\ -e^{i\Phi_0}\mathcal{Y}_o & 0 \end{pmatrix}. \qquad (10)$$

Then, using the "state Jones vector" $|\psi\rangle = \begin{pmatrix} \alpha \\ \beta \end{pmatrix}$ of the incident beam, and the corresponding state vector of the transmitted beam, $|\psi'\rangle = \hat{M}_0|\psi\rangle = \begin{pmatrix} e^{-i\Phi_0/2}\alpha \\ e^{i\Phi_0/2}\beta \end{pmatrix}$ ($\langle\psi'|\psi'\rangle = \langle\psi|\psi\rangle = 1$), we calculate the $x$- and $y$-shifts of the transmitted-beam centroid as the expectation values of operators (10):

$$\langle X \rangle = \langle\psi'|\hat{X}|\psi'\rangle = \tau\frac{1}{2k}\frac{d\Phi_0}{d\vartheta}, \qquad (11)$$

$$\langle Y \rangle = \langle\psi'|\hat{Y}|\psi'\rangle = \frac{\cot\vartheta}{k}\left[-\sigma(1-\cos\Phi_0) + \chi\sin\Phi_0\right]. \qquad (12)$$

Here we used Eqs. (6) and introduced the normalized Stokes parameters of the incident-beam polarization:

$$\tau = |\alpha|^2 - |\beta|^2, \quad \chi = 2\operatorname{Re}(\alpha^*\beta), \quad \sigma = 2\operatorname{Im}(\alpha^*\beta). \tag{13}$$

These parameters describe the degrees of extraordinary/ordinary, diagonal $\pm 45°$ linear, and right-/left-hand circular polarizations, respectively.

The polarization-dependent beam shifts $\langle X \rangle$ and $\langle Y \rangle$ are counterparts of the GH and IF shifts in the total internal reflection from an isotropic dielectric interface [24]. First, the analogue of the GH shift, Eq. (11), describes the *usual linear birefringence* between the ordinary and extraordinary rays. Here it is written in the form of the Artmann formulae [24]. Such a phase-gradient form of the birefringence shift was recently employed for a fine weak-measurement detection of the "photons trajectories", i.e., streamlines of the optical momentum density [42,43]. Note that the experiment [42] was set such that $\Phi_0 = 0 \mod 2\pi$ and the transverse effect in Eqs. (6) and (12) vanished: $\mathcal{Y}_{e,o} = \langle Y \rangle = 0$.

Second, the transverse shift (12) is the anisotropic counterpart of the IF shift or the spin Hall effect of light, and it is the central subject of our study. This transverse shift shows a new *transverse birefringence* of a uniaxial crystal plate, which is now caused by the finite size of the beam and is proportional to the $-\sigma(1-\cos\Phi_0) + \chi\sin\Phi_0$ polarization parameter. For $\Phi_0 = \pi \mod 2\pi$ this becomes purely *circular* birefringence, i.e., pure *spin-Hall effect* of light. Remarkably, this effect occurs already in the simplest anisotropic wave plates routinely used in optical setups but now *tilted* with respect to the beam propagation (for the normal incidence, the optical axis is orthogonal to the $z$-axis, $\vartheta = \pi/2$, and the effect vanishes: $\mathcal{Y}_{o,e} = \langle Y \rangle = 0$). Note also that, akin to the IF shift at interfaces, the spin-Hall-effect terms in the uniaxial crystal diverge for the propagation along the optical axis: $\vartheta \to 0$, $\cot\vartheta \to \infty$. This implies a singular transition to the cylindrically-symmetric problem of the on-axis propagation in uniaxial crystals: equation $\phi \simeq \Theta_y/\sin\vartheta$ is valid only in the $|\phi| \ll 1$ approximation. In the on-axis propagation, the SOI manifests itself as the spin-to-orbital angular-momentum conversion [25,26].

It is worth noticing that in the problem under consideration, the operators $\hat{X}$ and $\hat{Y}$, Eqs. (10), are Hermitian. Therefore, their expectation values (11) and (12) are purely real, which corresponds to the presence of spatial beam shifts and the absence of *angular* beam shifts (i.e., changes in the *direction* of the beam propagation) [24,38,40].

The spin-Hall effect can be measured either directly, via subwavelength shift (12) of the beam centroid [20,21,23,24], or via various other methods including *quantum weak measurements* [22,29–41]. The latter method allows significant amplification of the shift using almost crossed polarizers at the input and output of the system. As before, the input polarizer and matrix $\hat{M}_0$ determine the "pre-selected" polarization state $|\psi'\rangle = \hat{M}_0 |\psi\rangle = \begin{pmatrix} e^{-i\Phi_0/2}\alpha \\ e^{i\Phi_0/2}\beta \end{pmatrix}$, while the output polarizer corresponds to another, "post-selected" polarization state $|\varphi\rangle = \begin{pmatrix} \tilde{\alpha} \\ \tilde{\beta} \end{pmatrix}$. The resulting beam shifts after the second polarizer are determined by the *weak values* (instead of expectation values (11) and (12)) of the operators $\hat{\mathbf{R}}$, Eqs. (10). In contrast to the real expectation values of Hermitian operators, their weak values are *complex*. The real and imaginary parts of the weak values determine the spatial and angular beam shifts, $\langle \mathbf{R} \rangle_{\text{weak}}$ and $\langle \Theta \rangle_{\text{weak}}$, respectively [37–40]:

$$\langle \mathbf{R} \rangle_{\text{weak}} = \operatorname{Re}\frac{\langle\varphi|\hat{R}|\psi'\rangle}{\langle\varphi|\psi'\rangle}, \quad \langle \Theta \rangle_{\text{weak}} = \frac{1}{z_R}\operatorname{Im}\frac{\langle\varphi|\hat{R}|\psi'\rangle}{\langle\varphi|\psi'\rangle}. \tag{14}$$

As the beam propagates from its waist $z = 0$ along the $z$-axis, the angular shifts produce shifts in the beam centroid growing with $z$. The resulting shifts at $z \neq 0$ are

$$\langle \mathbf{R}_z \rangle_{\text{weak}} = \langle \mathbf{R} \rangle_{\text{weak}} + z \langle \mathbf{\Theta} \rangle_{\text{weak}} . \tag{15}$$

Thus, the quantum-weak-measurements technique can significantly amplify the beam shifts in two ways. First, the spatial shifts $\langle \mathbf{R} \rangle_{\text{weak}}$ can be much larger than the expectation values $\langle \mathbf{R} \rangle$ when $|\langle \varphi | \psi' \rangle| \ll 1$. Second, the appearance of angular shifts $\langle \mathbf{\Theta} \rangle_{\text{weak}}$, Eqs. (14), results in large beam shifts (15) in the far-field region: $z \gg z_R$.

Amplification of the regular birefringence shift, $\langle X \rangle_{\text{weak}}$, was previously measured in [33,34], and this was the first experimental example illustrating the quantum weak measurements paradigm. Quantum weak measurements were also used for amplification of the spin Hall effect shifts in the beam refraction and reflection at isotropic interfaces [22,35–41]. Here we analyze the amplification of the new spin Hall effect shift, $\langle Y \rangle_{\text{weak}}$. Let the incident beam be $e$-polarized, $|\psi\rangle = \begin{pmatrix} 1 \\ 0 \end{pmatrix}$, while the post-selection polarizer is almost orthogonal: $|\varphi\rangle = \begin{pmatrix} \sin\varepsilon \\ \cos\varepsilon \end{pmatrix} \simeq \begin{pmatrix} \varepsilon \\ 1 \end{pmatrix}$, $|\varepsilon| \ll 1$. Then, using Eqs. (6) and (10), equations (14) and (15) yield

$$\langle Y_z \rangle_{\text{weak}} = \frac{1}{\varepsilon k} \sin\Phi_0 \cot\vartheta + \frac{z}{z_R} \frac{1}{\varepsilon k} (1 - \cos\Phi_0) \cot\vartheta . \tag{16}$$

Here the first (spatial) and second (angular) terms correspond to the imaginary and real parts of the $\mathcal{Y}_{o,e}$ quantities, or to the $\chi$- and $\sigma$-dependent contributions to the regular beam shift (12). Importantly, the second term becomes dominant in the far-field zone and is amplified for two reasons: because $|\varepsilon| \ll 1$ and $z \gg z_R$. Note that the singular limit $\varepsilon \to 0$ is regularized by the condition of applicability of the above weak-measurement equations: $1 \gg |\varepsilon| \gg (kw_0)^{-1}$ [31,32]. Thus, the maximal achievable beam shift at $|\varepsilon| \sim (kw_0)^{-1}$ is of the order of the beam width in the far field: $w_0 z / z_R$. For the ordinary input polarization $|\psi\rangle = \begin{pmatrix} 0 \\ 1 \end{pmatrix}$ and post-selection $|\varphi\rangle \simeq \begin{pmatrix} 1 \\ -\varepsilon \end{pmatrix}$, the weak-measurement transverse shift is given by Eq. (16) with the "minus" sign before the first (spatial) term.

We demonstrate the weak measurements of the spin-Hall shifts (16) in the next Section. Here, to illustrate the beam-shift behavior, we calculate the centroid shifts (11) and (12) for a typical tilted anisotropic plate. As an example, we consider a quartz plate with thickness $d = 1$ mm. The phase difference between the ordinary and extraordinary waves is given by

$$\Phi_0(\vartheta) = k \left[ n_o d_o(\vartheta) - \tilde{n}_e(\vartheta) d_e(\vartheta) \right] . \tag{17}$$

Here $n_o = 1.544$ is the refractive index for the ordinary wave, $\tilde{n}_e(\vartheta) = n_e n_o / \sqrt{n_e^2 \cos^2\vartheta + n_o^2 \sin^2\vartheta}$ is the refractive index for the extraordinary wave propagating at the angle $\vartheta$ to the optical axis, $n_e = \tilde{n}_e(\pi/2) = 1.553$, and the distances of propagation of the ordinary and extraordinary rays in the tilted plate are

$$d_e(\vartheta) = \frac{\tilde{n}_e(\vartheta) d}{\sqrt{\tilde{n}_e^2(\vartheta) - \cos^2\vartheta}} , \quad d_o(\vartheta) = \frac{n_o d}{\sqrt{n_o^2 - \cos^2\vartheta}} . \tag{18}$$

Using Eqs. (17) and (18), in Figure 3 we plot the linear-birefringence and spin-Hall shifts (11) and (12) as functions of the tilt angle $\vartheta$. One can see that the transverse shift $\langle Y \rangle$ due to the spin Hall effect reaches wavelength-order magnitude, typical for other spin-Hall systems in optics [20–24]. In contrast to the IF shift in the reflection/refraction problems, here the transverse shift $\langle Y \rangle$ as a function of $\vartheta$ displays two-scale behavior. Namely, the fast oscillations in Fig. 3b originate from $(1-\cos\Phi_0)$ term with the rapidly growing (or decreasing) function $\Phi_0(\vartheta)$ (see Fig. 5 below), whereas the slow envelope corresponds to the universal $\cot\vartheta$ factor in SOI terms.

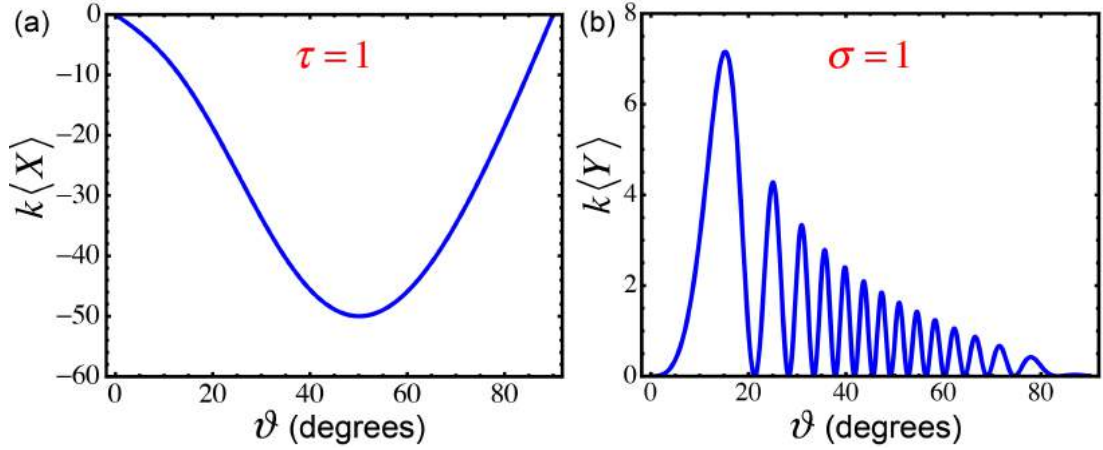

**Figure 3.** Longitudinal (in-plane) and transverse (out-of-plane) shifts of the beam transmitted through a tilted uniaxial crystal plate, Eqs. (11) and (12). These plots correspond to a 1 mm thick quartz plate and wavelength $\lambda = 2\pi/k = 632.8$ nm. The polarizations are: **(a)** $\tau = 1$ (extraordinary wave) and **(b)** $\sigma = -1$ (left-hand circular). For the opposite polarizations, $\tau = -1$ and $\sigma = 1$, the beam shifts have opposite sign, which signifies the usual in-plane linear birefringence between the ordinary and extraordinary waves, as well as the transverse circular birefringence, i.e., the spin Hall effect of light.

## 4. Experimental results

To verify the above theoretical predictions, we performed a series of experimental measurements using the setups shown in Figure 4. For the anisotropic plate, we used a multiple-order half-wave plate (WPMH05M-670, Thorlabs, USA) made of crystalline quartz and designed to operate at a wavelength of 670 nm. As a source of the incident Gaussian beam, we employed a semiconductor laser diode of wavelength $\lambda = 2\pi/k = 675$ nm. The laser radiation was passed through a single-mode fiber and collimated using a microscope objective lens.

Since the wavelength of the beam differed from the nominal wavelength of the wave plate, we chose to measure the anisotropic phase difference $\Phi_0$ versus the angle of the tilt $\vartheta$ via direct Stokes-polarimetry methods [28,44] instead of calculating it via Eqs. (17) and (18). For this purpose we used the setup shown in Fig. 4a. The double Glan-Thomson polarizer P1 selected the desired linear-polarization state in the incident beam. In the first experiment, this was 45° polarization, i.e., $\begin{pmatrix} \alpha \\ \beta \end{pmatrix} = \frac{1}{\sqrt{2}} \begin{pmatrix} 1 \\ 1 \end{pmatrix}$. Then, the beam passed through the tilted wave plate, a quarter-wave plate (QWP) with retardation angle $\delta$, and the second polarizer P2 with angle $\gamma$. We measured the integral intensity of the transmitted beam, $\bar{I}'(\delta,\gamma)$ ($\bar{I}' = \int I'(\mathbf{R})d^2\mathbf{R} = \int |\mathbf{E}'(\mathbf{R})|^2 d^2\mathbf{R}$), and determined the integral Stokes parameters in the transmitted beam as

$$\overline{S}_0 = \overline{I}'(0°,0°)+\overline{I}'(0°,90°), \quad \overline{S}_1 = \overline{I}'(0°,0°)-\overline{I}'(0°,90°),$$
$$\overline{S}_2 = \overline{I}'(0°,45°)-\overline{I}'(0°,135°), \quad \overline{S}_3 = \overline{I}'(90°,45°)-\overline{I}'(90°,135°). \quad (19)$$

(Note that the normalized Stokes parameters (13) $(\tau,\chi,\sigma)$ correspond to $\overline{S}_{1,2,3}/\overline{S}_0$ in the incident beam.) Finally, the phase difference between the ordinary and extraordinary modes was determined as [28,44]

$$\Phi_0 = \tan^{-1}\left(\frac{\overline{S}_3}{\overline{S}_2}\right). \quad (20)$$

The measured phase (20) versus the tilt angle $\vartheta$ of the anisotropic plate is shown in Figure 5.

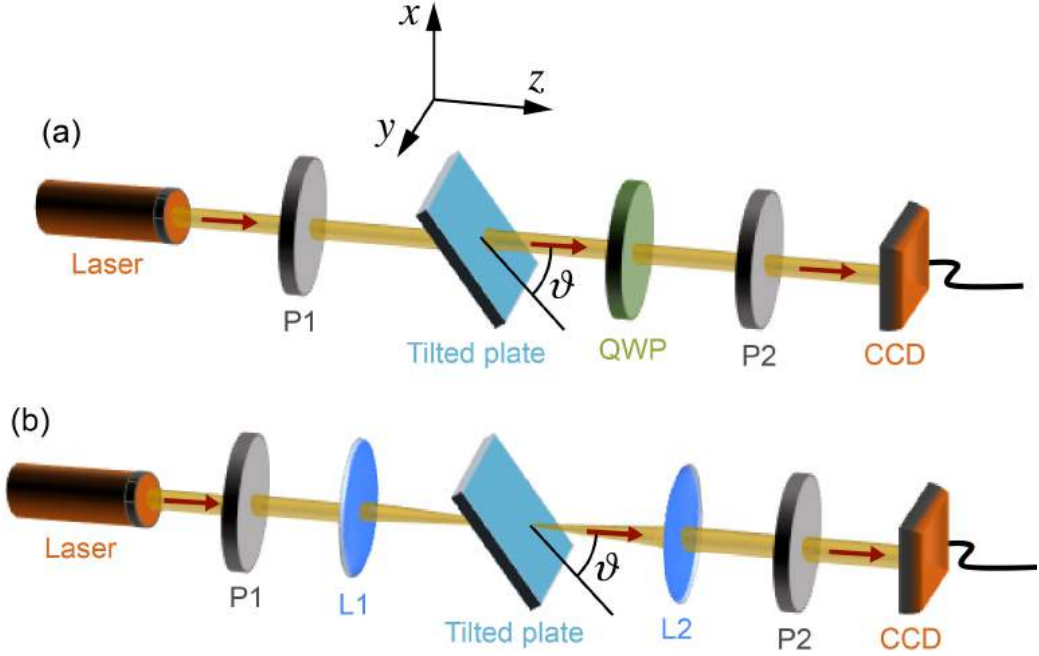

**Figure 4.** Schematics of the experimental setups used for the polarimetric measurements **(a)** and quantum weak measurements of the beam shifts **(b)**.

The phase difference $\Phi_0(\vartheta)$ calculated via the *integral* Stokes parameters (19) completely characterize the action of the tilted wave plate of the central plane wave in the beam. To investigate the spin-Hall effect in the transmitted beam, we preformed a series of measurements of *local* intensity distributions $I'(\mathbf{R})$ and corresponding local Stokes-parameters $S_i(\mathbf{R})$ and in the beam cross-section.

First, we measured the distributions of the normalized third Stokes parameter in the transmitted beam,

$$s_3(\mathbf{R}) = \frac{S_3(\mathbf{R})}{S_0(\mathbf{R})}, \quad (21)$$

which characterizes the local ellipticity of the field, or the normalized $z$-component of its spin angular momentum density [45]. These distributions are shown in Figure 6 for the extraordinary and ordinary polarizations of the incident beams. In agreement with theoretical predictions, Fig. 2, one can clearly see the transverse $y$-separation of positive and negative ellipticities (cf. [13,20]). This is the first experimental confirmation of the spin-Hall effect of light produced by the transmission through a tilted anisotropic plate.

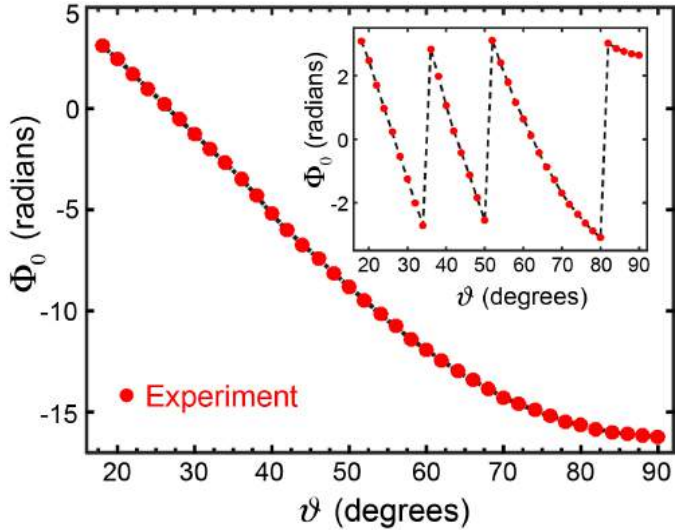

**Figure 5.** The polarimetrically-measured phase difference $\Phi_0(\vartheta)$, Eq. (20), between the ordinary and extraordinary modes for the beam transmitted through the tilted half-wave plate. The inset shows the actual measured phase in the range $(-\pi,\pi)$, whereas the main plot shows the unwrapped phase.

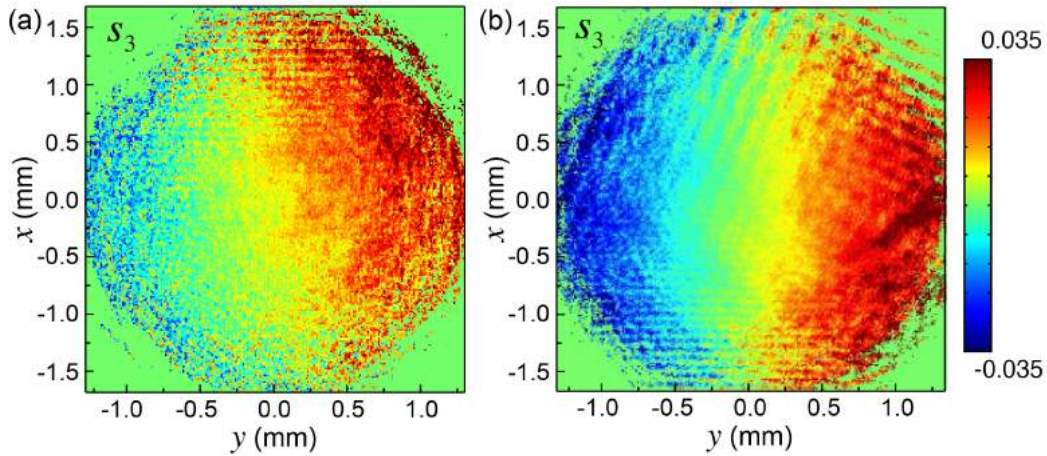

**Figure 6.** Experimentally-measured distributions of the local Stokes parameter $s_3(\mathbf{R})$, Eq. (21), in the extraordinary **(a)** and ordinary **(b)** beams transmitted through the tilted half-wave plate with $\vartheta \simeq 35°$ and $\Phi_0(\vartheta) \simeq -\pi$ (see Fig. 5). The $y$-splitting of opposite spin states with $s_3 > 0$ and $s_3 < 0$ corresponds to the splitting of opposite polarization ellipticities in Fig. 2 and signals the spin-Hall effect of light.

Second, we investigated deformations of the intensity distributions and beam shifts using the quantum weak-measurement method described in Section 3. We used the setup shown in Fig. 4b with additional focusing (L1) and imaging (L2) lenses. The two polarizers P1 and P2 produced pre-selected and post-selected polarization states $|\psi\rangle$ and $|\varphi\rangle$, respectively, while the lenses controlled the amplification propagation factor $z/z_R$ in Eq. (16). Namely, the first lens L1 of focal length 5 cm produced a focused Gaussian beam with the Rayleigh range $z_R \simeq 3840\,\mu\text{m}$ (determined from the $1/e^2$ spot size of the original laser beam, 374 μm, and the spot size of the focused beam in the focal plane, 28.72 μm), while the second lens L2 of focal length 5 cm collimated the beam and provided the effective propagation distance $z = 5$ cm. Thus, the propagation amplification factor was $z/z_R \simeq 13$, and the second, angular term in the beam shift (16) strongly dominated the first, spatial term (cf. [22,23]).

We performed weak-measurement experiments with the pre-selection in the $e$-polarized state, $|\psi\rangle = \begin{pmatrix} 1 \\ 0 \end{pmatrix}$, and post-selection in the almost-orthogonal state $|\varphi\rangle = \begin{pmatrix} \sin\varepsilon \\ \cos\varepsilon \end{pmatrix} \simeq \begin{pmatrix} \varepsilon \\ 1 \end{pmatrix}$, $|\varepsilon| \ll 1$, as well as with the pre-selection in the $o$-polarized state $|\psi\rangle = \begin{pmatrix} 0 \\ 1 \end{pmatrix}$ and post-selection in $|\varphi\rangle \simeq \begin{pmatrix} 1 \\ -\varepsilon \end{pmatrix}$. In both cases, the transverse beam shift is described by the second (angular) term in Eq. (16). The transverse intensity distributions $I'(\mathbf{R})$ in the $o$-polarized beam transmitted through the tilted half-wave plate and post-selected with $\varepsilon = -1.4 \cdot 10^{-2}, 0, 1.4 \cdot 10^{-2}$ are shown in Figure 7. One can clearly see beam deformations typical for quantum weak measurements [30–34]. Namely, the two-hump Hermitte-Gaussian $y$-distribution takes place for $\varepsilon = 0$, whereas Gaussian-like distributions are considerably shifted in opposite $y$-directions for $\varepsilon = \pm 1.4 \cdot 10^{-2}$. These intensity deformations and shifts provide the second experimental evidence of the transverse circular birefringence and spin-Hall effect in the system.

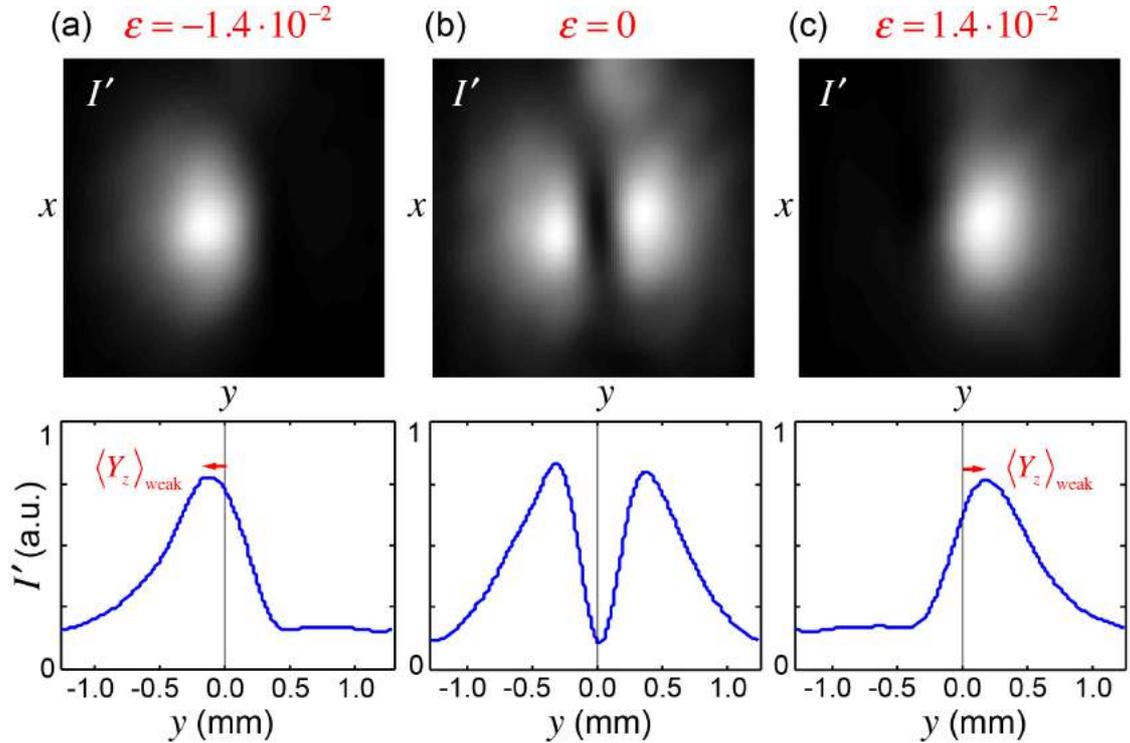

**Figure 7.** The transverse intensity distributions $I'(\mathbf{R})$ in the $o$-polarized beam transmitted through the tilted half-wave plate and post-selected in the almost $e$-polarized state with $\varepsilon = -1.4 \cdot 10^{-2}, 0, 1.4 \cdot 10^{-2}$ (see explanations in the text). The two-hump Hermitte-Gaussian distribution at $\varepsilon = 0$ corresponds to Eq. (8), while the opposite shifts $\langle Y_z \rangle_{\text{weak}}$ at $\varepsilon = \pm 1.4 \cdot 10^{-2}$ are the spin-Hall shifts amplified via quantum weak measurements, Eq. (16). Like in Fig. 6, the tilt angle with $\vartheta \simeq 35°$ corresponds to $\Phi_0(\vartheta) \simeq -\pi$, i.e., maximizes the spin-Hall effect $\propto (1 - \cos\Phi_0)$.

The transverse $y$-shifts of the Gaussian distributions in Fig. 7 are the beam shifts $\langle Y_z \rangle_{\text{weak}}$ described by Eq. (16). These are strongly amplified from the typical subwavelength scale $k^{-1}$, Eq. (12), to the beam-width scale with the overall weak-measurement amplification factor

$$A = \frac{1}{|\varepsilon|} \frac{z}{z_R} \simeq 929 \ . \tag{22}$$

The experimentally-measured transverse beam shift $\langle Y_z \rangle_{weak}$ versus the tilt angle $\vartheta$ are plotted in Figure 8 for the *e* and *o* pre-selected polarizations and the corresponding post-selections with $\varepsilon = 1.4 \cdot 10^{-2}$. Since the phase difference $\Phi_0(\vartheta)$ is known from independent polarimetric measurements (Fig. 5), we compare the measured beam shifts with the analytical result in Eq. (16). Figure 8 shows a very good agreement between the experiment and theory. This provides the quantitative confirmation of the spin-Hall effect and circular birefringence of light transmitted through a tilted anisotropic plate.

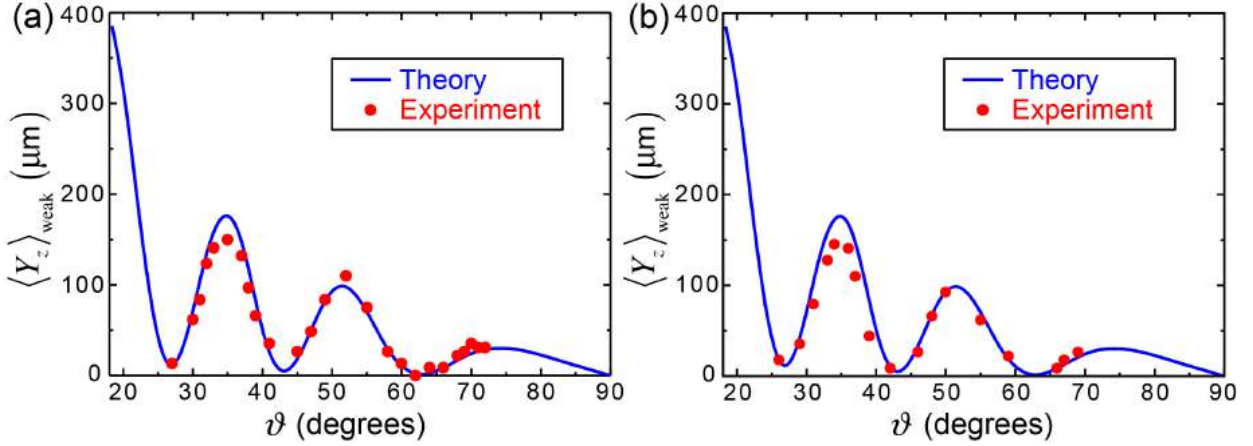

**Figure 8.** Transverse beam shifts $\langle Y_z \rangle_{weak}$ determined via quantum weak measurements (Fig. 7) and theoretically calculated using Eq. (16) with the phase $\Phi_0(\vartheta)$ taken from polarimetric measurements (Fig. 5). The measurements are done for the *e*-polarized **(a)** and *o*-polarized **(b)** input beams, and the post-selection with $\varepsilon = 1.4 \cdot 10^{-2}$ (see explanations in the text).

## 5. Conclusions

We have considered the transmission of a Gaussian light beam through a uniaxial crystal plate with a tilted anisotropy axis. The action of the plate on a plane wave is well-known and is described by the diagonal Jones matrix with a phase retardation between the ordinary and extraordinary polarizations. However, birefringence phenomena require the consideration of confined beams rather than infinite plane waves. We have shown that taking into account multiple plane waves with slightly different wavevector directions in the beam spectrum results in nontrivial beam shift effects. First, the transmitted beam experiences the in-plane shift between the *o* and *e* linear polarizations. This is the well-known *linear* birefringence. Second, the beam experiences a transverse out-of-plane shift dependent on the circular (and also diagonal) polarization degrees, i.e., a *circular* (and diagonal) birefringence. This is a manifestation of the spin-orbit interaction and a novel type of the spin-Hall effect of light.

Notably, the usual linear birefringence and new circular birefringence form a close analogy with the Goos–Hänchen and Imbert–Fedorov beam shifts that appear in the light reflection at a dielectric interface. This is because mathematically-similar spin-orbit interactions appear: (i) in the beam reflection due to the medium *inhomogeneity* and different Fresnel coefficients for the TE and TM polarizations and (ii) in the beam transmission through a crystal plate due to the medium *anisotropy* and different transmission coefficients for the *o* and *e* polarizations.

We have provided a detailed theoretical description and experimental measurements of the novel circular-birefringence phenomenon. The remarkably simple system of a tilted half-wave plate and polarizers was used for this. Our measurements clearly demonstrated the spin-Hall effect and transverse beam shifts in the transmitted beam via both polarimetric and quantum-weak-measurement methods. Using the weak-measurement technique we strongly enhanced the transverse beam shift to the beam-width size and also transformed the *spatial* shift into an *angular* shift, which is clearly seen in the far field.

Thus, we have described a novel basic phenomenon in a simple thoroughly-studied system. Due to the great recent interest in optical spin-orbit interaction phenomena and the wide use of anisotropic plates in numerous optical setups and devices, our results could find applications in polarization optics and nano-photonics. The methods developed in this work can be extended and applied to other types of anisotropic plates: dichroic, circular-birefringent, etc.

## Acknowledgements

We acknowledge support from RIKEN iTHES Project, MURI Center for Dynamic Magneto-Optics via the AFOSR (grant number FA9550-14-1-0040), Grant-in-Aid for Scientific Research (A), the Australian Research Council, Department of Science and Technology (DST), India, University Grants Commission – Basic Science Research (UGC-BSR) fellowship, D.S. Kothari post-doctoral fellowship, PICT-2014-1543, PICT-2015-0715, and UBACyT-PDE-2015.